\documentclass[fleqn,usenatbib]{mnras}

\usepackage{newtxtext,newtxmath}

\usepackage[T1]{fontenc}
\DeclareRobustCommand{\VAN}[3]{#2}
\let\VANthebibliography\thebibliography
\def\thebibliography{\DeclareRobustCommand{\VAN}[3]{##3}\VANthebibliography}

\usepackage{algorithm}
\usepackage{lscape}
\usepackage{amsmath}	% Advanced maths commands
\usepackage{float}
\restylefloat{table}
\usepackage[varg]{txfonts}

\newcommand{\rom}[1]{\uppercase\expandafter{\romannumeral #1\relax}}
\newcounter{mylabelcounter}

\makeatletter
\newcommand{\labelText}[2]{%
#1\refstepcounter{mylabelcounter}%
\immediate\write\@auxout{%
  \string\newlabel{#2}{{1}{\thepage}{{\unexpanded{#1}}}{mylabelcounter.\number\value{mylabelcounter}}{}}%
}%
}
\makeatother

\defcitealias{toba14}{1}
\defcitealias{dong12}{2}
\defcitealias{LH11}{3}
\defcitealias{crook07}{4}
\defcitealias{shirazi12}{5}
\defcitealias{shim13}{6}
\defcitealias{allen11}{7}
\defcitealias{liu11}{8}
\defcitealias{oh15}{9}
\defcitealias{zhou06}{10}
\defcitealias{brinchmann08}{11}

\usepackage{gensymb}
\usepackage{ragged2e}
\usepackage{lscape}
\usepackage{booktabs}
\usepackage{natbib}
\bibpunct{(}{)}{;}{a}{}{,}
\usepackage{supertabular}
\usepackage{array}
\usepackage{float}
\usepackage{varioref}
\usepackage{multirow}
\usepackage{graphicx}
\usepackage{appendix}
\usepackage{epstopdf}
\usepackage{longtable}
\usepackage{comment}
\usepackage{threeparttable, tablefootnote}
\usepackage[T1]{fontenc}
\usepackage{etoolbox}
\usepackage{siunitx}
\usepackage{caption}
\usepackage{subcaption}
\usepackage{tabu}
\usepackage{tikz-cd}
\usepackage{ulem}
\DeclareGraphicsExtensions{.eps}

\title[Constraints on extragalactic transmitters]{\LARGE Constraints on extragalactic transmitters via Breakthrough Listen observations of background sources}
\author[M.A. Garrett \& A.P.V. Siemion]{M.A. Garrett$^{1, 2}$
and A.P.V. Siemion$^{3, 4, 1}$
\\
$^{1}$Jodrell Bank Centre for Astrophysics (JBCA), Department of Physics \& Astronomy, Alan Turing Building, The University of Manchester, M13 9PL, UK\\
$^{2}$Leiden Observatory, Leiden University, PO Box 9513, 2300 RA Leiden, The Netherlands\\
$^{3}$Berkeley SETI Research Center, University of California Berkeley, Berkeley CA 94720, USA\\
$^{4}$SETI Institute, Mountain View, CA 94043, USA\\
}

\date{Submitted 14 March 2022 / Revised 2 September 2022 / Accepted 2 September 2022}
\pubyear{2022}
\begin{document}
\label{firstpage}
\pagerange{\pageref{firstpage}--\pageref{lastpage}}
\maketitle
%%%%%%%%%%%%%%%%%%%%%%%%%%%%%%%%%%%%%%%%%%%%%%%%%%%%%%%%%%%%%%%--ABSTRACT--
\begin{abstract}
%Context

The Breakthrough Listen Initiative has embarked on a comprehensive SETI survey of nearby stars in the Milky Way that is vastly superior to previous efforts as measured by a wide range of different metrics. SETI surveys traditionally ignore the fact that they are sensitive to many background objects, in addition to the foreground target star. In order to better appreciate and exploit the presence of extragalactic objects in the field of view, the Aladin sky atlas and NED were employed to make a rudimentary census of extragalactic objects that were serendipitously observed with the 100-m Greenbank telescope observing at 1.1-1.9~GHz. For 469 target fields (assuming a FWHM radial field-of-view of 4.2 arcminutes), NED identified a grand total of 143024 extragalactic objects, including various astrophysical exotica e.g. AGN of various type, radio galaxies, interacting galaxies, and one confirmed gravitational lens system. Several nearby galaxies, galaxy groups and galaxy clusters are identified, permitting the parameter space probed by SETI surveys to be significantly extended. 
Constraints are placed on the luminosity function of potential extraterrestrial transmitters assuming it follows a simple power-law and limits on the prevalence of very powerful extraterrestrial transmitters associated with these vast stellar systems are also determined.  It is demonstrated that the recent Breakthrough Listen Initiative, and indeed many previous SETI radio surveys, place stronger limits on the prevalence of extraterrestrial intelligence in the distant Universe than is often fully appreciated.

\end{abstract}

%%%%%%%%%%%%%%%%%%%%%%%%%%%%%%%%%%%%%%%%%%%%%%%%%%%%%%%%%%%%%%%--1.Intro--
\section{Introduction} \label{1}

The Breakthrough Listen Initiative \citep{Worden2017} is conducting the most thorough and comprehensive SETI (Search for Extraterrestrial Intelligence) programme ever conducted. The initiative greatly surpasses previous efforts in all regards but particularly in terms of spectral and temporal resolution, data processing techniques and compute capacity, observing time, raw sensitivity and human effort. The results of the initial first phase of observations have been published \citep{enriquez2017, Price2020, vishal2021, raffy2021, franz2022} and the first signals of interest have also been identified and rigourously analysed \citep{sheikh2021, smith2021}. The longer term goal of the Breakthrough Listen Initiative over the next few years is to survey one million nearby stars, the entire Galactic plane, the galactic centre and a diverse range of 123 nearby galaxies located at distances of up to 30~Mpc. 

The initial searches have focused on targeting stars in the Milky Way, in particular nearby systems. Recognising that these radio observations are typically also sensitive to many other cosmic objects in the radio telescope's natural field-of-view, \citet{Bart2020} extended the Breakthrough Listen (hereafter BL) surveys of 1327 stars targeted by \citet{enriquez2017} and \citet{Price2020} to take into account other stellar objects in the field. In particular, stars with parallaxes measured by {\textit{Gaia}}, extended the sample from the original 1327 target stars to 288315 stellar objects, thus permitting \citet{Bart2020} to place much tighter limits on the prevalence of nearby extraterrestrial transmitters. 

Most recently, \citet{Tremblay2022} have used the Murchison
Widefield Array (MWA) at $\sim 150$~MHz to perform a blind search of more that 3 million stars toward the Galactic
Centre but with relatively course spectral resolution ($\sim 10$~kHz).    

%SETI searches that traditionally neglect the presence of background sources in the field of view do not do full justice to the data, and miss the opportunity to extend the SETI analysis to a great number and wide range of nearby and distant extragalactic phenomena.

What is also not always fully appreciated is that radio SETI observations targeting nearby stars, have a field of view that naturally includes a great number and wide range of different background \textit{extragalactic} systems. Indeed, this point is traditionally ignored in the analysis of SETI data. However, this fact is of significant interest because although extragalactic systems are located at distances that are many orders of magnitude larger than the nearest galactic stars, they contain hundreds of billions of stars, all simultaneously located within the main response of the radio telescope beam. Assuming these galaxies are not unlike our own Milky Way, a significant fraction of these stars may host planets located within the habitable zone \citep{Batalha12647}. And while the luminosity function of extraterrestrial transmitters is currently unknown, it might not be too surprising if it was represented by a power-law suggesting that a population of rare but very powerful sources might be observable at great distance (e.g. \citet{Drake1973}). 

\citet{Gray2017} performed a very well motivated search of two nearby galaxies - M31 and M33 at a distance of 760 and 970 kpc respectively, using the JVLA. Multiple pointings were used in order to cover the full optical extent of both galaxies. Although the spectral resolution and bandwidth of this observation was relatively modest, the large number of stars observed simultaneously made this a rather pertinent search strategy - one that is very relevant to the work presented in this paper.     

%($$\sim 10^{12}$, demonstrated the   

In a very welcome effort to expand the diversity of
targets surveyed by the BL Initiative, \citet{lacki2021} has compiled an additional catalogue of possible targets, with the ambition of including "one of everything" - a complete range of cosmic phenomena or astrophysical "exotica". The catalogue includes 816 distinct targets and is remarkably wide ranging, including minor bodies of the solar system, planetoids, giant planets, stars, collapsed objects, stellar groups, nebulae and the ISM, galaxies, active galactic nuclei (AGN), galaxy associations, and large-scale structures.  In addition to comprising a large amount of baryonic matter in diverse configurations, this sample represents environments that may very well be preferred for technologically capable life that may be much more evolved than our own species and/or have evolved along exotic trajectories. 

The main thrust of the research presented here is to consider the type of extragalactic objects (and indeed astrophysical exotica) that are routinely observed by SETI radio surveys that target nearby stars but unavoidably also include other background sources within the telescope's natural field of view. In this paper, an attempt has been made to assess the impact of this effect by making a study of a 469 field subset of the 692 fields first observed by \citet{enriquez2017} using the 100-m Greenbank Telescope (GBT) at L-band. The form of the paper is as follows - in section \ref{2} the field selection is considered and the approach used to identify extragalactic objects also described. Section \ref{3} presents the main results, including some detailed examples of the exotica present in these fields. Section \ref{4} uses these results to expand the parameter space explored by SETI surveys, and to place limits on the prevalence of very powerful extraterrestrial transmitters. Section \ref{5} presents the overall conclusions of the paper. In what follows, a flat cosmology with $\rm H_{o} \text{ = } 70$~km/s/Mpc and $\Omega_{\rm M} \text{ = }0.3$ is adopted.

%%%%%%%%%%%%%%%%%%%%%%%%%%%%%%%%%%%%%%%%%%%%%%%%%%%%%%%%%%%%%%%--2.The LOFAR DR1 Value-added Catalogue--
\section{Field Selection and object identification} \label{2}

\subsection{Field Selection} \label{2.1}

In support of the BL observing programme, \citet{isaacson2017} created an initial target list of 1709 nearby stars, including a range of different spectral type. The sample includes the 60 nearest stars ($< 5$~pc) with the remaining stars located within $5-50$~pc. \citet{enriquez2017} observed 692 of these 1709 stars with the 100-m Greenbank Telescope (GBT) at 1.1-1.9~GHz. 

Not surprisingly, many of the BL stars are also located near the galactic plane. These fields can be very crowded, and the chances of misidentifying galactic stars as extragalactic objects is significant. In order to reduce potential contamination by galactic sources, 92 fields were removed from the sample of 692 stars for which the target star was located within $\pm 10$ degrees of the galactic plane. Since the stars observed are typically nearby, many of them are also quite bright, especially in the optical. The very bright diffraction patterns associated with these stars makes it difficult to reliably identify other sources in the rest of the field. In order to alleviate this effect, stars with $m_{V} < 5$ were identified and a further 120 fields were removed from the reduced sample. 
Finally, because many of the target stars are actually located in multiple stellar systems, 11 cases of overlapping fields were identified, and for each pair, one field was removed from the sample. This left a grand total of 469 fields in the final sample. 

\subsection{Object identification} \label{2.2}

Following \citet{Bart2020}, the FWHM of the GBT at L-band was adopted to be 8.4 arcminutes. All object searches were therefore conducted employing cone searches with a radius of 4.2 arcminutes. At the edge of the cone, the sensitivity of the GBT is reduced by a factor of 2, compared to on-axis observations. It should therefore be noted that this radial cutoff is therefore somewhat arbitrary since the GBT retains some sensitivity beyond this point. 

\begin{figure}
\includegraphics[width=\linewidth]{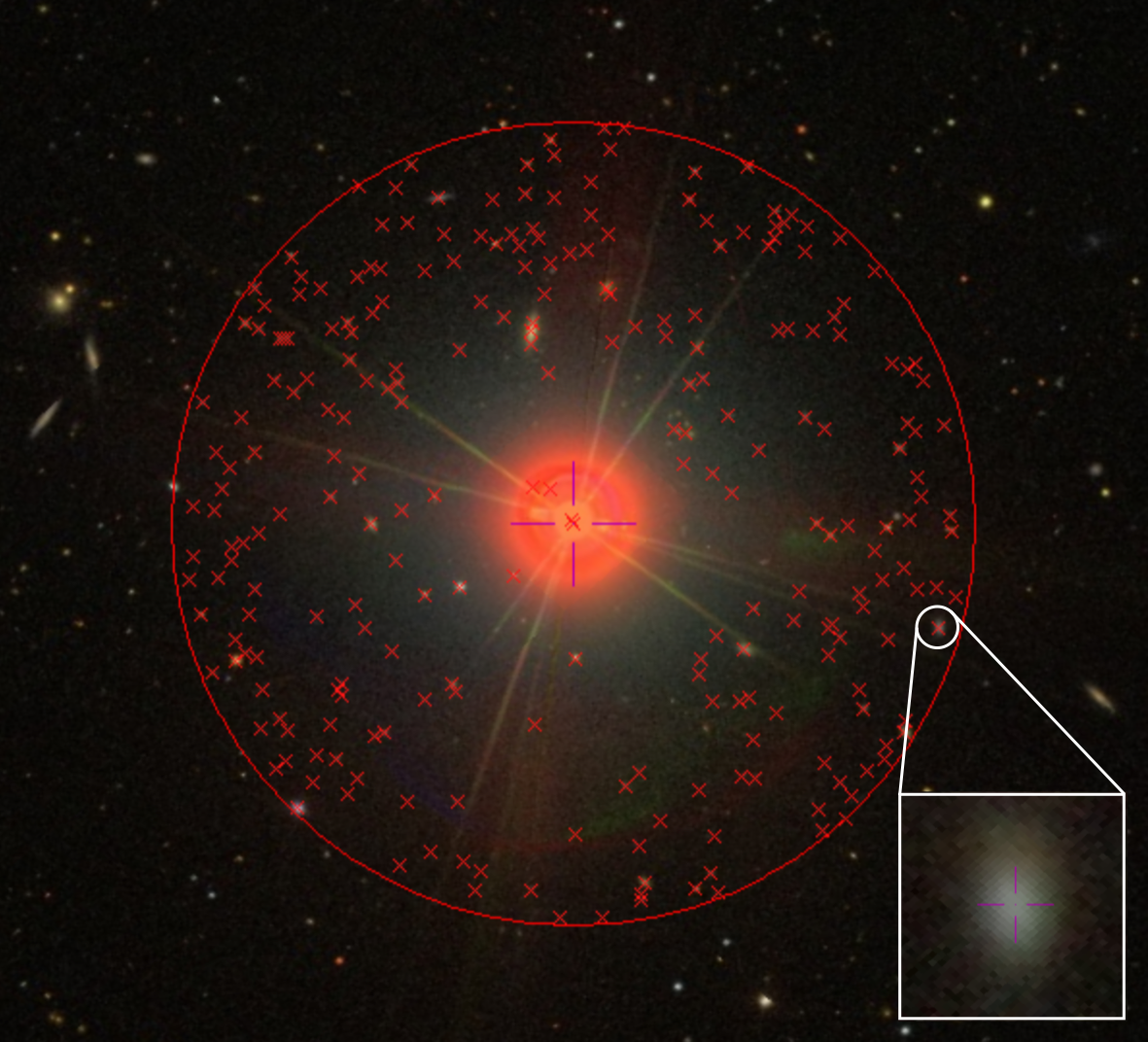}
\caption{An optical SDSS9 colour image of the stellar field centred on  BL~J033901.1-053736.5, showing the extent of the FWHM for the GBT at L-band circled in red. The NED identifications in the field are represented by red crosses. The nearby galaxy, WISEA J033845.71-053842.1 is circled in the main image, with a close up view presented at the bottom right of the figure. The scale of the square inset is $15 \times 15$ arcseconds.
} 
\label{f1}
\end{figure}

The NASA/IPAC Extragalactic Database ({\tt\string NED}\footnote{\url{https://ned.ipac.caltech.edu/}}) was used to search for objects within the central 4.2 arcminutes of each field. {\tt NED} is the largest and most comprehensive online database of cosmic objects that lie beyond the Milky Way \citep{Mazzarell2020}. For each of the 469 fields, a cone search of radius 4.2 arcminutes was performed. Some additional information on the fields was obtained by repeating the search on each field using  {\tt\string SIMBAD}\footnote{\url{http://simbad.u-strasbg.fr/simbad/}} \citep{Wenger2000}. Extensive use of {\tt\string Aladin} \footnote{\url{https://aladin.cds.unistra.fr/}} \citep{Bonnarel2000, Mazzarell2020} was made to interrogate both the NED and SIMBAD data base servers, and to generate multi-wavelength images of each field. In particular, images were made for each field with Aladin  via the the DSS2, PanSTARSS DR1, SDSS9, 2MASS, GALEX, AllWISE, WENSS and NVSS survey collections. The Aladin Macro controller was used to drive scripts that greatly reduced the administrative burden of executing these repetitive tasks on each of the 469 fields. It should be noted that the centre of each field was defined as the pointing position used by the GBT. Since some of the stars have large proper motions, the multi-wavelength images (typically made at a range of different epochs) often show the central star (or stellar system) offset from the centre of the field. Fig.\ref{f1} presents a typical field.

%%%%%%%%%%%%%%%%%%%%%%%%%%%%%%%%%%%%%%%%%%%%%%%%%%%%%%%%%%%%%%--3.Results--
\section{Results} 
\label{3}

A total of 143024 objects were identified  by {\tt NED}, including 17810 point sources, 28405 galaxies, 87841 Infrared sources, 44 QSOs, 8016 Ultraviolet sources, 401 X-ray sources, 398 radio sources, 11 Absorption line systems, 5 Gamma ray sources, 53 Galaxy cluster members, 33 galaxy groups, 6 galaxy pairs and 1 galaxy triple. For this sample, {\tt NED} also reports spectroscopic redshifts for 989 objects and photometric redshifts for 434 objects. Clearly many of the objects fall into the category of "astrophysical exotica" as defined earlier by \citet{lacki2021}. In the following sub-sections, some examples of the wide range of various exotica are presented, with an emphasis on those systems that are relatively nearby and for which some constraints on the nature and prevalence of powerful transmitters may be inferred. 

\subsection{Selected exotica}\label{3.1}

\subsubsection{Nearby galaxies} \label{3.1.1}

Visual inspection of the SDSS and DSS2 optical images of each field, reveals a significant number of extended, nearby galaxies contained within the 4.2 arcminute FWHM of the GBT. For sources identified by NED as ordinary galaxies, 6 are convincing examples of galaxy scale systems with $z < 0.02$. Table \ref{T1} presents the main properties of these systems. Fig.\ref{f1} includes WISEA J033845.71-053842.1 -  a galaxy located at $z\text{ = }0.010824$.

\subsubsection{Interacting galaxies and star forming regions} \label{3.1.2}

\begin{figure}
\includegraphics[width=\linewidth]{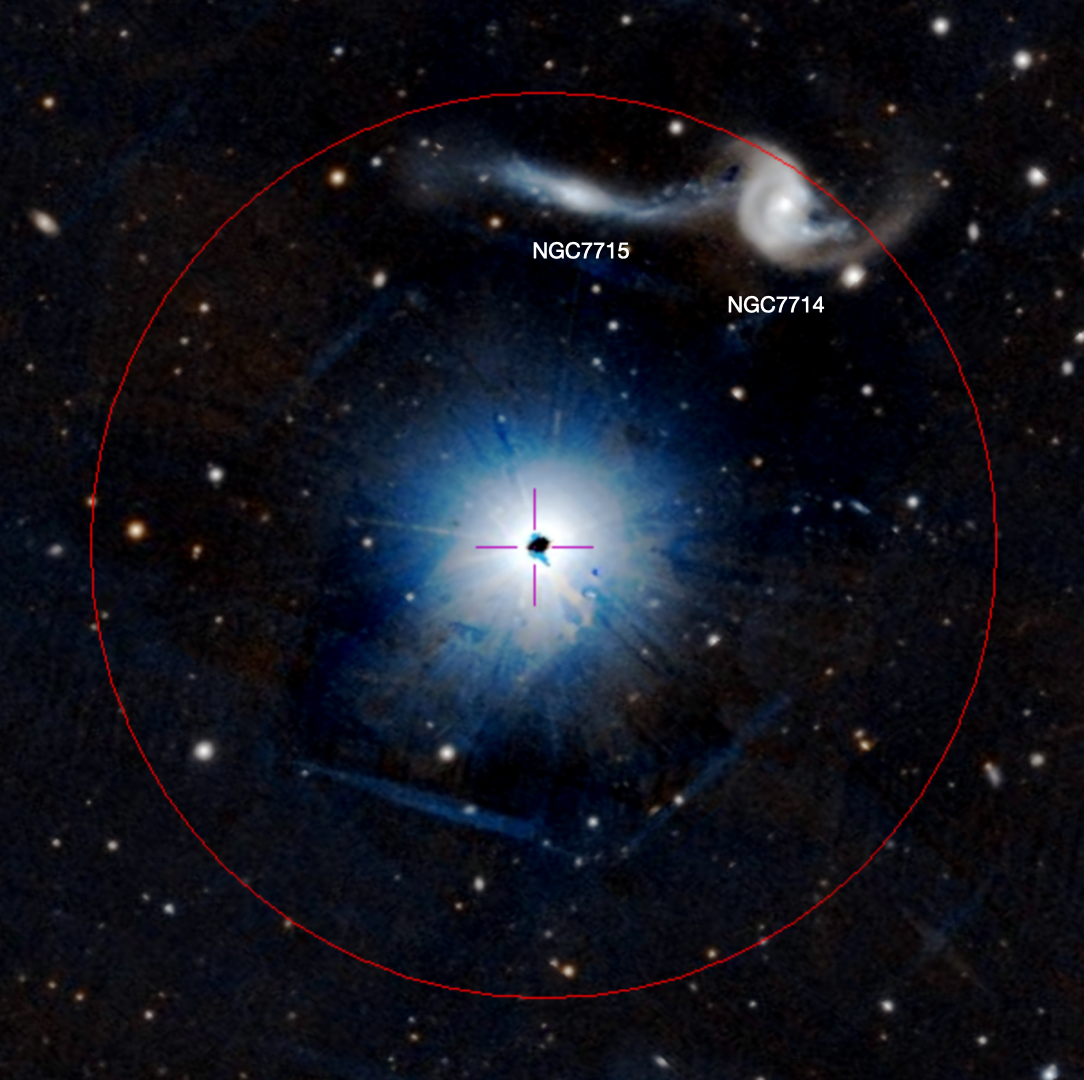}
\caption{An optical PanSTARRS DR1 z and g broadband filters colour image of the stellar field centred on  BL~J233623.1+020609.0, showing the extent of the FWHM for the GBT at L-band circled in red. The interacting galaxies NGC~7714 and NGC~7715 are clearly labeled and are located close to the edge of 4.2 arcminute search cone.} 
\label{f2}
\end{figure}

The field BL~J233623.1+020609.0 provides a good example of an interacting pair of galaxies - NGC 7714 at $z \text{ = } 0.009333$ and NGC 7715 at $z \text{ = } 0.009243$. NGC7714 (also known as Arp 284) is a prototypical starburst galaxy that is interacting with its smaller neighbour NGC 7715 - the average star formation rate is $\sim 1$M$_{\odot}y^{-1}$ and its properties are comparable to M82 (see \citet{Lancon2001} and references therein). NGC 7714 is detected across the electromagnetic spectrum, and is also associated with NVSS~J233614+020918, 2 supernova events (SN 1999dn \& SN 2007fo) and 2 possible Ultra-luminous X-ray candidates. 
Located at a distance of $\sim 40$~Mpc this system is only a little further away than the BL galaxy sample's limit ($ < 30$~Mpc). Fig.\ref{f2} shows the field BL~J233623.1+020609.0, containing NGC 7714 and NGC 7715. 

\subsubsection{Galaxy clusters} \label{3.1.3}

The presence of nearby galaxy clusters (and galaxy groups) is of special interest since it greatly increases the number of potential extraterrestrial transmitters simultaneously present within a single field of view, located at the same distance (see also section \ref{4.2}). The nearest galaxy cluster is located in the field BL J143328.4+525435.3 - there are eight galaxies with measured cluster redshifts in the GBT's 4.2 arcminutes radial field of view  - Table \ref{T2} presents the main properties of the cluster members.

%WHL~J143313.0+525748 ($z \text{ = } 0.046300$), SDSS J143342.88+525346.9 ($z \text{ = } 0.045790$), WISEA J143343.08+525350.4 ($z \text{ = }0.045783$), WISEA J143342.42+525320.3 ($z \text{ = }0.042923$), [DRS2017] 000192 ($z \text{ = }0.045000$), WISEA J143325.87+525053.5 ($z \text{ = }0.047088$),  
%Mr19:[BFW2006] 20354    ($z \text{ = }0.046650$)  and 
%MCG +09-24-014 ($z \text{ = } 0.047299$). 

\begin{figure}
\includegraphics[width=\linewidth]{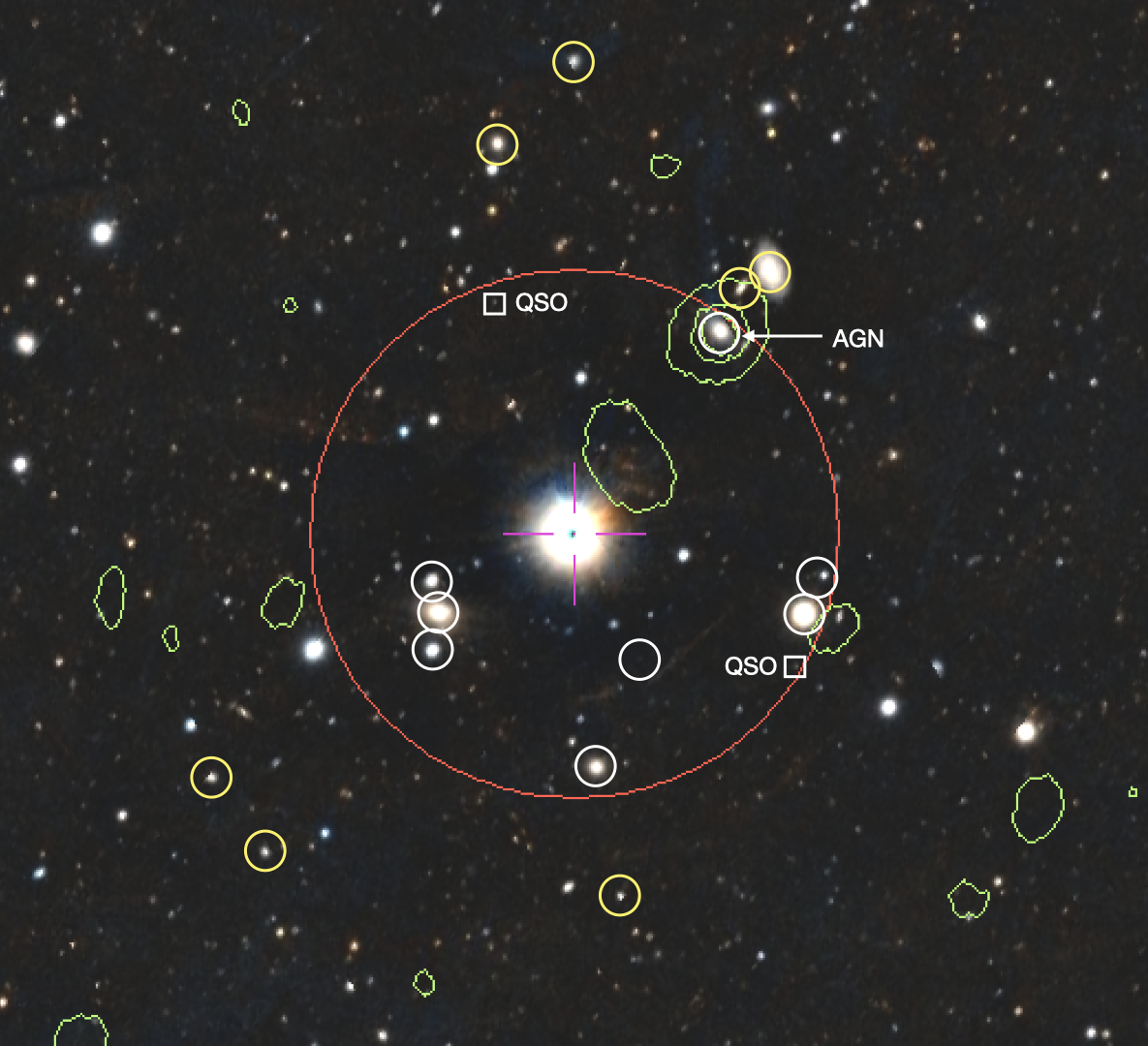}
\caption{An optical PanSTARRS DR1 z and g broadband filters colour image of the stellar field centred on  BL J143328.4+525435.3, showing the extent of the FWHM for the GBT at L-band circled in red. Members of the cluster located at $z \sim 0.046$ are identified with the white circles. Several prominent members of this cluster lying outside of the 4.2 arcminutes FWHM are also identified with yellow circles. Several AGN are also located within the search cone, including 2 QSOs (labelled) and MCG +09-24-014 which is an AGN.} 
\label{f3}
\end{figure}

Fig.~\ref{f3} shows the field BL J143328.4+525435.3. It should be noted that this field also contains several AGN located at higher redshift. 

\subsubsection{Nearby AGN} \label{3.1.4}

The nearest AGN identified by NED is WISEA~J080816.59+210857.0 (also identified as SDSS J080816.59+210857.1) located at a redshift of $z \text{ = } 0.142090$ in field BL~J080812.8+210612.2. The source is included in the quasars and active nuclei catalogue of \citet{Veron2010}. \citet{Kyuseok2015} classify the source as a low-luminosity Type 1 AGN  in which the central engine is viewed directly without significant obscuration.

\subsubsection{Radio-loud galaxies} \label{3.1.5}

Somewhat ironically, a few of the BL fields contain relatively bright radio loud sources detected in various surveys, including NVSS \citep{Condon1998} and WENSS \citep{Wenss1997}. Prominent examples in our sample include: NVSS 132105+341502 and
NVSS~J194137+503431 (3C304). The main properties of NVSS~132105+341502 (see \citep{Caccianiga2008} and \citet{Veron2010}) are presented in Table \ref{T3}.  

%NVSS 132105+341502 has an integrated flux density of $474\pm17$~mJy at 1.4 GHz with a deconvolved size $< 17.2$~arcseconds. It appears to also be identified with the radio sources 87GB~131847.5+343057, B2~1318+34B and 6C B131847.2+343044, and is detected at IR, Optical and X-ray wavelengths. 
%(WISEA~J132105.56+341500.9 and 2CXO~J132105.4+341459)
%It is a radio-loud AGN with a spectroscopic redshift of $z\text{ = } 0.452$ \citep{Caccianiga2008}, and is included in the quasars and active nuclei catalogue of \citet{Veron2010}.  

\begin{figure}
\includegraphics[width=\linewidth]{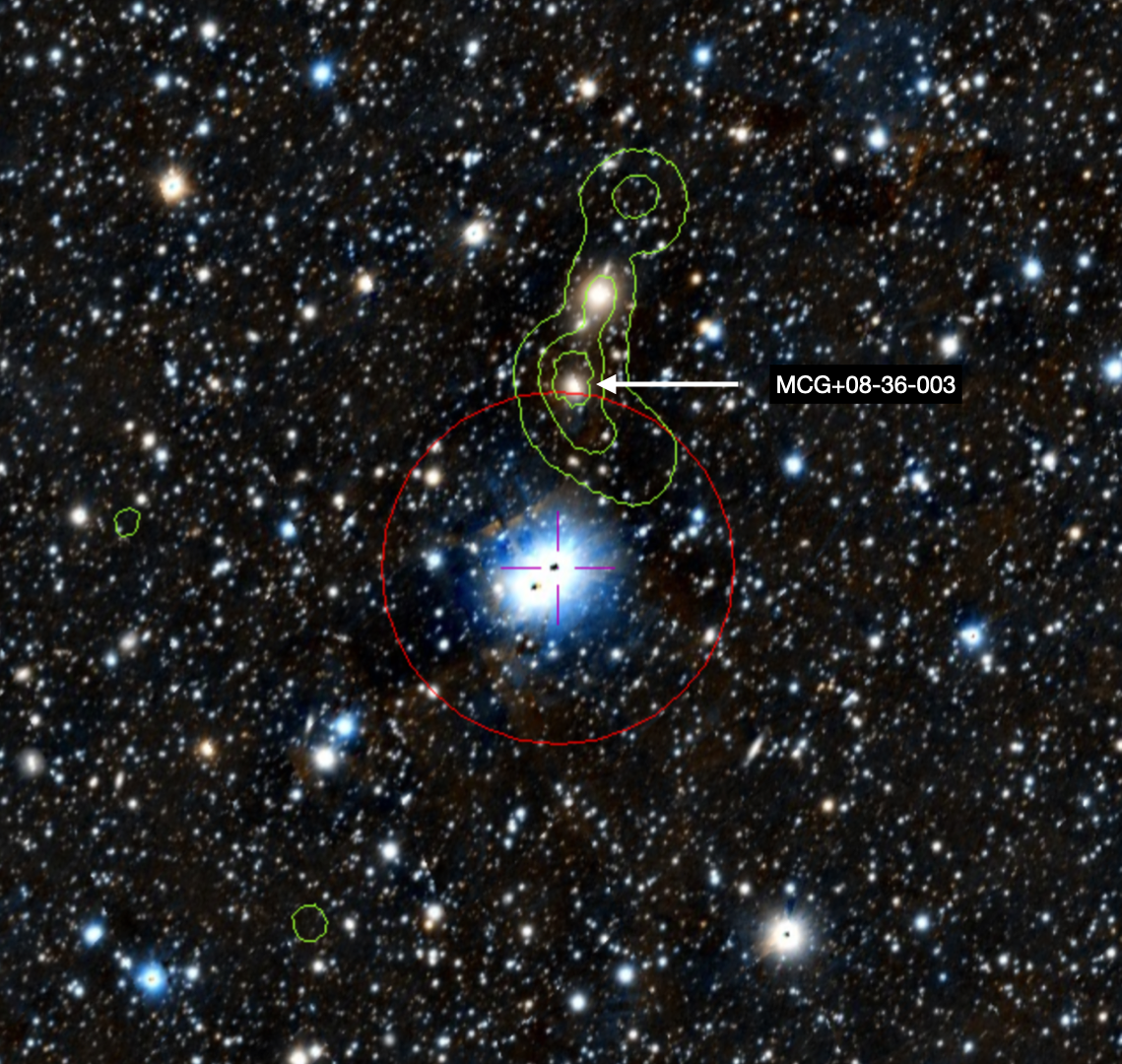}
\caption{An optical PanSTARRS DR1 z and g broadband filters colour image of the stellar field centred on  J194148.6+503127.5, showing the extent of the FWHM for the GBT at L-band circled in red. A contour plot of 3C402 is superimposed upon the optical image. MCG+08-36-003 is identified in the figure, straddling the 4.2 arcminute search cone.} 
\label{f4}
\end{figure}

 NVSS~J194137+503431 is a radio source that straddles the 4.2 arcminute search cone of BL field J194148.6+503127.5, and is part of the very bright and extended radio source 3C402. 3C402 has an integrated flux density of $\sim 10.1\pm1.0$~Jy at 178MHz \citep{Kellermann1969} and is associated with  NVSS J194136+504017 ($0.586
\pm 0.019$~Jy), NVSS J194146+503548               $(1.884 \pm 0.058)$~Jy,  and NVSS J194142+503750 $0.489\pm0.015$~Jy, in addition to NVSS J194137+503431               $0.151\pm0.005$~Jy. Fig.\ref{f4} presents 3C402 superimposed upon a PanSTARRS optical image of the field BL~J194148.6+503127.5. Recent highlights from the on-going VLASS survey, resolve 3C402 into at least two separate radio galaxies \citep{Lacy2020}. 3C402 (NVSS~J194137+503431) is identified with a galaxy group - the details are presented in Table \ref{T4}. 

%nearby low surface brightness galaxy MCG+08-36-003 (with further cross-identifications including CGCG 257-007) with $M_{V}\text{ = }14.0$ and a spectroscopic redshift $z\text{ = }0.02595$. In the optical, MCG+08-36-003 straddles the 4.2 arcminute search cone, and appears to be part of a group of galaxies that includes UGC 11465 ($z\text{ = }0.026061$), WISEA J194140.93+503832.5 ($z\text{ = }0.027592$) and 2MASX J19414453+5037151 ($z\text{ = }0.027805$).     
 
 %(6C B194024.0+502903)
 
 %MCG+08-36-003 i while UGC 11465 lies well outside of the search area.  

%NVSS~J194136+504017 VLSS J1941.6+5040 - furthers north radio emission.

%CGCG 257-007 z=0.025948 3c402

%6C B194024.0+502903
%MG4 J194131+5039
%NVSS J194136+504017
%NVSS J194137+503431
%MG4 J194202+5035

% contours are drawn at 0.01, 0.1 and 0.35 Jy/beam. 

%MCG+08-36-003 z = distance 114 Mpc

%NVSS J161710+551746

%13:20:58.6|+34:16:39.4
%

\subsubsection{Gravitational lenses} \label{3.1.6}

One confirmed gravitational lens system is identified, MM184222+593828 \citep{Lestrade2010}, and one lens candidate, CFHTLS J020832.1-043315 \citep{Anupreeta2016}. Our focus lies with the confirmed system since this also has a measured redshift.    
MM184222+593828 was first identified as a high redshift Sub-mm Galaxy using the MAMBO bolometer array at IRAM \citep{Lestrade2010}. Initially,  extremely large lensed magnifications ($\mu$) were implied ($\mu \sim 300$) in order to account for estimated IR luminosity of the system but radio observations by \citet{McKean2011} using the Westerbork Synthesis Radio Telescope (WSRT) at 1.4~GHz suggested much more modest values were required. VLA observations of CO \citep{Lestrade2011} determined the redshift of the system to be $z \text{ = }3.93$ and revealed a complete Einstein ring with a diameter of 1.4 arcseconds. A revised estimate of the IR luminosity of the system by \citet{Lestrade2011}, was consistent with the analysis of \citet{McKean2011}, implying a lens magnification of $\mu \sim 12$ \citep{Lestrade2011}. 

%CFHTLS J020832.1−043315 \citep{Anupreeta2016} was discovered as part of the citizen scientist SPACE WARPS lens search, utilising the Canada–
%France–Hawaii Telescope Legacy Survey (CFHTLS). This lens system appears to be a double image system with one of the images stretched into an arc with an Einstein radius of 1.4 arcseconds. The redshift of the images is currently unknown. The system appears to be an example of a galaxy scale lens with a photometric redshift of $z \sim 1$ a. The lens mass is identified as an isolated  red starforming galaxy. Some modeling of this and other SPACE WARPS systems have been made \citep{Kung2018} but no value for the image magnification is reported.   

\section{Placing limits on the prevalence of distant extraterrestrial transmitters} 
\label{4}

\subsection{SETI survey figure of merit} 
\label{4.1}

\citet{enriquez2017} first introduced the Continuous
Waveform Transmitter Figure of Merit (CWTFM) for SETI surveys. The CWTFM reflects the likelihood of finding an extraterrestrial signal above a specific minimum Equivalent Isotropic Radiated Power ($\rm{EIRP_{min}}$), such that: 

\begin{equation}
    {\rm CWTFM} \text{ = } \zeta_{AO}~\frac{{\rm EIRP_{min}}}{N_{\rm *}\nu_{\rm rel}},
\end{equation}

where $N_{\rm *}$ is the total number of stars observed per pointing and the fractional bandwidth $\nu_{\rm rel}$ is the total bandwidth of the receiver normalized by the central observing frequency, and $\zeta_{\rm AO}$ is a normalising constant such that CWTFM is 1 for an EIRP equal to that of former Arecibo Planetary Radar system ($10^{13}$W). Smaller values for the CWTFM implies surveys that are more complete and/or have better sensitivity. For example, the initial survey of \cite{enriquez2017} has a CWTFM $\sim 0.85$. \citet{Price2020} made a significant improvement on the results of \cite{enriquez2017}, by employing a larger range of Doppler drift rate ($\pm 4$~Hz~s$^{-1}$) and lowering the SNR of the original analysis from $20$ to $> 10$, and including more observations with GBT and Parkes radio telescopes at S-band, achieving a CWTFM of $\sim 0.11$. By including other galactic stars in the telescope's field of view, \citet{Bart2020} improved further upon Price's figure, obtaining a CWTFM of 0.04 for nearby stars. For comparison, the previous extragalactic search made by \citet{Gray2017} has a CWTFM of $\sim$ 136. 

% \zeta_{AO}=5E-11. 

% w=1.699*FWHM/2 
% I/Io=exp(-2x^{2}/w^{2})
% I/Io=exp(-0.0393*X^2)

% EIRP=2.1E12*((DDDD E6/50)**2)*RRR

% TR = 1/(NNN*(0.66/1.5)

% CWTFM= EIRP (5E-11*RRR*2.1E12*((DDDD E6/50)**2))/(NNN E11*0.8/1.5)= 
% CWTFM= 5E-11*EIRP/(NNN*0.8/1.5)

% eg Enriquez 5E-11*5.2E12/(692*(0.66/1.5))

% this work: 5E-11*EIRP/(NNN*(0.66/1.5))

%Tremblay is 1.1E19 W EIRP 1/TR = 1/(3E6*(30.72E6/154.4E6))=1.67E-6 = -5.77

Following \citet{Bart2020}, the EIRP$_{\rm min}$ obtained for distant objects in the field of view is estimated by scaling from the values already determined by \citet{Price2020} - at a distance of 50~pc, the EIRP$_{\rm min}$ for the GBT at L-band is $2.1\times10^{12}$~W. By adopting a Gaussian function for the telescope beam shape, the reduced response of the telescope at the position of the source is also taken into account. For simplicity, it is assumed that the galaxies considered here contain about the same number of stars ($N_{*}$)  as the Milky Way,  more specifically we add a level of uncertainty to this figure, assuming $N_{*} \sim 10^{11\pm0.3}$. We believe that this is a reasonable assumption - in particular, the objects used to derive the results presented later in Fig.~\ref{f5}, are all massive galaxies with absolute magnitudes that are comparable with or in excess of the Milky Way. 

\subsection{Results} 
\label{4.2} 

Table \ref{T5} presents the CWTFM, EIRP$_{\rm min}$ and Transmission Rate ($\rm TR$) values for some of the astrophysical exotica discussed in the previous section. The results suggest that interesting figures of merit can be achieved for these systems, especially nearby galaxy groups and clusters - in these cases the very large number of stars that are simultaneously in the telescope field of view partially compensates for the loss in sensitivity (a larger EIRP$\rm{_{min}}$). It's interesting to include the gravitational lens MM184222+593828 in this analysis because of the large source magnification factor ($\sim 12$) but the EIRP$_{min}$  for this system remains extremely large. 

\begin{figure*}
\includegraphics[width=\linewidth]{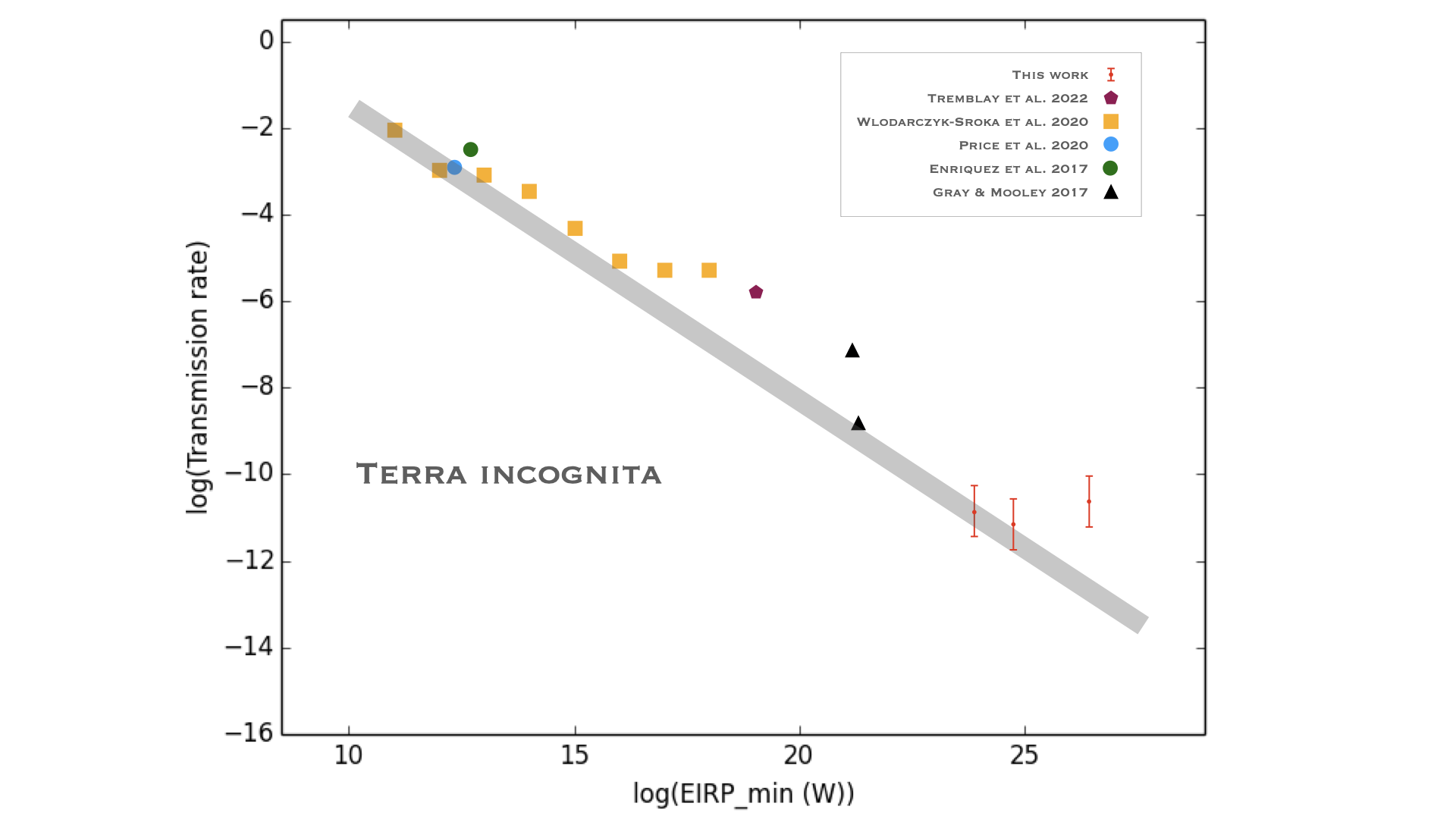}
\caption{A plot of the Transmission Rate against $\rm EIRP_{min}$ for a range of SETI studies. The gray line is defined by the two most constraining values - this work (the results for NGC~7714/NGC~7715, the 3C402 galaxy group, and the nearby AGN WISEA~J080816.59+210857.0) and that of \citet{Bart2020} - these data extend the parameter space into a new region of smaller transmission rates and larger $\rm EIRP_{min}$. If extraterrestrial transmitters follow a power-law distribution, the region below this line represents a "terra incognita" for SETI surveys. The error bars of this work (red symbols) are presented at the 2-sigma level for clarity. } 
\label{f5}
\end{figure*}

Following \citet{enriquez2017}, \citet{Price2020} and \citet{franz2022}, a subset of the results presented here are compared with previous surveys - in particular, Fig.\ref{f5} plots the "transmission rate", $\rm TR$, ($\frac{{1}}{N_{\rm *}\nu_{\rm rel}}$) against $\rm EIRP_{min}$ for various surveys. Note that the error bars of this work (red symbols) are determined by the uncertainty in the value of $N_{*}$ associated with the nearby galaxies and galaxy groups identified in the field - for clarity these are presented at the 2-sigma level. Inclusion of our results for the interacting pair NGC~7714/NGC~7715, the 3C402 galaxy group, and the nearby AGN WISEA~J080816.59+210857.0 extend the parameter space into a new region of smaller transmission rates and larger $\rm {EIRP_{min}}$. If extraterrestrial transmitters follow a power-law distribution as shown in Fig.\ref{f5}, they are probably confined to a region below the line that is a best fit to the results of \citet{Bart2020} and the new analysis presented here. This region represents "terra incognita" for SETI studies and its full exploration requires surveys with better sensitivity and a wider field of view. Fig.\ref{f5} also shows two gaps in $\rm EIRP_{min}$ coverage - the first is due to the physical distance between the most distant stars in the Milky Way and our nearest galactic neighbours, and the second is due to the physical distance that separates our nearest galactic neighbours and the more distant galaxies considered here ($z \sim 0.01-0.14$). Targeting nearby galaxy groups and clusters, could address the latter issue while going deeper with longer integration times, would seem to be another promising approach. 

 It may be wise to sound a note of caution at this point with respect to whether adopting a simple power-law distribution for extraterrestrial transmitters is appropriate. Astronomers, and in particular radio astronomers, are very familiar with power-laws that operate over many orders of magnitude but there is no evidence to suggest that the luminosity function for extraterrestrial transmitters will be similarly broad, especially on the scales considered here. Indeed while some authors have followed \citet{Drake1973} in assuming a simple power-law distribution for extraterrestrial transmitters (e.g. \citet{Gulkis1984} and \citet{Lampton2002}), others have rejected this assumption entirely, (e.g. \citet{Dreher2004}), arguing that delta functions may be more appropriate for engineered technology. A less extreme approach would be to adopt more nuanced probability functions, such as the Pareto distribution but without knowing the scales on which the distribution flattens, their application is also limited for our purposes. Even if a power-law is indeed appropriate, the slope could be significantly steeper than that shown in Fig.\ref{f5}. 

Since the BL observations have an integration time of typically $3 \times  5$ minutes per target field, these surveys and the implied prevalence of powerful transmitters only relates to high-duty cycle transmitters that are in prolonged and continuous operation. With this as a caveat, we note that \citet{enriquez2017} and \citet{Price2020} find no evidence for 100\% duty cycle transmitters in the original survey, and it is therefore also possible to place some limits on the prevalence of such transmitters at extragalactic distances. 

For example, in the case of NGC 7714 and 7715, no evidence is found for transmitters with an EIRP in excess of $7.5\times10^{23}$~W. We define the prevalence of extraterrestrial transmitters as the fraction of stars in this interacting system that host a powerful ($\rm EIRP > 7.5\times10^{23}$~W) long-duty cycle transmitter at the time the observations were made. For this particular system, the prevalence is $< 10^{-12}$.  

% https://uncertaintycalculator.com/
% prevalence = 1/(2*10^(11±0.5))
% TR = -log((2*10^(11±0.3))*(0.66/1.5))
% CWTFM = ((5*10^-11)*(7.5*10^23))/((2*10^(11±0.3)*0.8)/1.5)

We note that the gain of an antenna similar in scale to the Square Kilometre Array (SKA) is $\sim 10^8$, so the intrinsic transmitter power levels required are of order $10^{16}$~W, {\it i.e.} a total energy supply similar to that associated with a Kardashev Type I civilisation \citep{kardashev64} but only a tiny fraction of that available to a Type II civilisation ($\sim 10^{26}$~W). These power levels are many orders of magnitude greater than the most powerful radio transmitters routinely in use on Earth today ($ EIRP \sim 10^{13}$~W), and it seems appropriate to consider whether such systems could be realised in practice. Currently, our most powerful radar systems combine together relatively modest transmitter powers with large phased arrays {\it e.g} Space Force. However, as noted earlier by \citet{isaacson2017}, there are no known physical laws that would preclude much larger EIRP figures to be attained 
and as receiver technology approaches the quantum noise limit, this is the only way to extend the range of many communication and radar applications. 
\citet{Benford2010a} have considered favourably the feasibility of constructing messaging beacons operating at microwave frequencies with EIRPs approaching $\sim 10^{20}$~W using existing technology. These installations employ phased arrays of thousands of powerful transmitters distributed on scales of up to $\sim 20$~km$^{2}$ with the associated costs of construction estimated to be $\sim \$ $20 billion. Other sources of such powerful transmissions include operational installations {\it e.g.} microwave beamers for interstellar sails \citep{Benford2019}. It's difficult to predict what civilisations more advanced than our own might be capable of but surely it would not be too surprising if they employed much more powerful transmitters, assuming they routinely operate on interstellar scales.  
%track, propel and communicate with their own spacecraft routinely operating on vast inter-stellar spatial scales.

%Similar power levels have been considered for beaming technology  However as receiver technology approaches its theoretical limits, range expansion will rely on increasing raw transmitter power or antenna gain. This will be particularly true for deep space communications.  It's difficult to predict what very advanced civilisations might be capable of, but enormous space-borne phased arrays tied to powerful transmitters is surely not beyond the grasp of Kardashev Type I and II civilisations. Such arrays would be required in order to track and communicate with free-flying megastructures or interstellar outposts deployed on vast spatial scales and could also be used as "beacons" for civilisations wishing to signal their existence and attract the attention of other, perhaps less advanced but emerging technical civilisations such as our own.   

%this power requirement is only a technological hurdle, not an obstacle of physical principle. - ISaacson et al. 

% calculateD Dl via NED calculator 

%%%%%%%%%%%%%%%%%%%%%%%%%%%%%%%%%%%%%%%%%%%%%%%%%%%%%%%%%%%%%%%--3.Results--

%%%%%%%%%%%%%%%%%%%%%%%%%%%%%%%%%%%%%%%%%%%%%%%%%%%%%%%%%%%%%%%--5.Conclusions--
\section{Summary} 
\label{5}

The current Breakthrough Listen programme, employing large single dish radio telescopes, also observes many other interesting sources in the field-of-view, in addition to the main target star. This paper has focused on the background sources that fill the field of view, including a wide range of varied astrophysical "exotica". These include nearby galaxies, galaxy groups, galaxy clusters, distant gravitational lens systems and a broad range of different AGN type. Extending the original BL analysis to take into account the potential response of extraterrestrial transmitters in these very distant systems, permits us to extend the parameter space explored by previous surveys and to place new constraints on the luminosity function of extraterrestrial transmitters, assuming the latter follow a simple power-law description. 

Future work might choose to focus on regions of parameter space that represent SETI's "terra incognita" - suggesting that future surveys should consider: (i) going much deeper by employing longer integration times and/or more sensitive instruments (e.g. the SKA telescope now under construction) and (ii) targeting as large a number of potential targets simultaneously in a single pointing (e.g. nearby galaxies and galaxy groups/clusters). Combining these two approaches is another promising line of study e.g. long "stares" towards nearby groups and clusters. Finally, we note that SETI surveys that use beam-forming techniques to generate a limited number of relatively narrow pencil beams, will reduce the fraction of sky that is accessible to serendipitous discovery, as further detailed in \cite{drake00}. Employing interferometric techniques on antenna arrays that maintain the full field-of-view of the individual dishes would offer a better capability in this regard but at the price of significantly larger data volumes. 

%%%%%%%%%%%%%%%%%%%%%%%%%%%%%%%%%%%%%%%%%%%%%%%%%%%%%%%%%%%%%%%--Acknowledgments--
\section*{Acknowledgements}

This research has made use of (i) the NASA/IPAC Extragalactic Database (NED),
which is operated by the Jet Propulsion Laboratory, California Institute of Technology,
under contract with the National Aeronautics and Space Administration, (ii) the SIMBAD database,
operated at CDS, Strasbourg, France and (iii) the "Aladin sky atlas" also developed at CDS, Strasbourg Observatory, France. 

We thank the referee for comments that helped us improve upon the original submitted paper, and we acknowledge discussions with James Benford on his work on the feasibility of powerful beacon transmitters.  

\section*{Data Availability}
The data underlying this article are available in the article.
%%%%%%%%%%%%%%%%%%%%%%%%%%%%%%%%%%%%%%%%%%%%%%%%%%%%%%%%%%%%%%%--References--
\bibliographystyle{mnras} % style aa.bst
\bibliography{ref}

%%%%%%%%%%%%%%%%%%%%%%%%%%%%%%%%%%%%%%%%%%%
%%%%%%%%%%%%%%%%%%%%%%%%%%%%%%%%%%%%%%%%%%%
%%%%%%%%%%%%%%%%%%%%%%%%%%%%%%%%%%%%%%%%%%%
%%%%%%%%%%%%%%%%%%%%%%%%%%%%%%%%%%%%%%%%%%%
%%%%%%%%%%%%%%%%%%%%%%%%%%%%%%%%%%%%%%%%%%%
%%%%%%%%%%%%%%%%%%%%%%%%%%%%%%%%%%%%%%%%%%%

%%%%%%%%%%%%%%%%%%%%%%%%%%%%%%%%%%%%%%%%%%%%%%%

\clearpage
\onecolumn
\landscape 

%These include: WISEA J033845.71-053842.1 ($z\text{ = }0.010824$ and associated with Breakthrough Listen (BL) field BL~J033901.1-053736.5), WISEA J132337.71+291716.9 (a $z\text{ = }0.013501$ starburst galaxy, [SMB88] 2617), WISEA J132814.15-021828.8 ($z\text{ = }0.012270$, also identified  as 2dFGRS TGN197Z212), NGC 5797 ($z\text{ = }0.013269$), IC~4563 (a $z\text{ = }0.014258$ liner) and NGC~5966 ($z\text{ = }0.015048$). IC~4563 and NGC~5966 are located in the same field, BL~J153555.9+394952.0 (see Fig.\ref{f1}). 

\vspace{1cm}

\begin{center}

\begin{longtable}{lll}

\caption{The properties of nearby galaxies contained within various BL fields (see section \ref{3.1.1})}

\label{T1} \\
\makebox[\textwidth]{
    \begin{tabular}{lllcc}
        \toprule
        \multirow{3}{*}{BL Field name} & \multirow{3}{*}{Source name} & \multirow{3}{*}{Type } & \multirow{3}{*}{$z$ } & \multirow{3}{*}{Other ID }  \\
        & & & &  \\
        \midrule
        BL~J033901.1-053736.5 & WISEA J132337.71+291716.9 & Starburst galaxy & 0.013501 & [SMB88] 2617 \\
        BL~J132821.2-022143.9 & WISEA J132814.15-021828.8 & Galaxy &  0.012270 & 2dFGRS TGN197Z212  \\
        BL~J145622.9+493744.4 & NGC 5797  & Galaxy & $0.013269$ & - \\
        BL~J153555.9+394952.0 & IC~4563 & Liner & 0.014258 & - \\
        BL~J153555.9+394952.0 & NGC~5966  & Galaxy Group & $0.015048$  & - \\
        \midrule
        \bottomrule
    \end{tabular}}
    
    \end{longtable}

\vspace{1cm}

%WHL~J143313.0+525748 ($z \text{ = } 0.046300$), SDSS J143342.88+525346.9 ($z \text{ = } 0.045790$), WISEA J143343.08+525350.4 ($z \text{ = }0.045783$), WISEA J143342.42+525320.3 ($z \text{ = }0.042923$), [DRS2017] 000192 ($z \text{ = }0.045000$), WISEA J143325.87+525053.5 ($z \text{ = }0.047088$),  
%Mr19:[BFW2006] 20354    ($z \text{ = }0.046650$)  and 
%MCG +09-24-014 ($z \text{ = } 0.047299$). 

\begin{longtable}{lclc}
\caption{The main properties of the cluster members associated with BL~J143328.4+525435.3 (see section \ref{3.1.3})}
\label{T2} \\
\makebox[\textwidth]{
    \begin{tabular}{lclc}
        \toprule
         \multirow{3}{*}{Source name} & \multirow{3}{*}{Type } & \multirow{3}{*}{$z$ } & \multirow{3}{*}{Other ID }  \\
        & & &  \\
        \midrule
        WHL~J143313.0+525748 & galaxy & 0.046300 & - \\
        DSS~J143342.88+525346.9 & galaxy &  $0.045790$ & -  \\
        WISEA J143343.08+525350.4 & galaxy & $0.045783$ & - \\
        WISEA J143342.42+525320.3 & galaxy & 0.042923 & - \\
        DRS2017 000192 & galaxy &  0.045000 & - \\
        WISEA J143325.87+525053.5 & galaxy & 0.047088 & - \\
        BFW2006 20354  & galaxy & 0.046650 & - \\
        MCG +09-24-014 & galaxy & 0.047299 & - \\
        
        \midrule
        \bottomrule
    \end{tabular}}
    \end{longtable}

\vspace{1cm}

\clearpage

%NVSS 132105+341502 has an integrated flux density of $474\pm17$~mJy at 1.4 GHz with a deconvolved size $< 17.2$~arcseconds. It appears to also be identified with the radio sources 87GB~131847.5+343057, B2~1318+34B and 6C B131847.2+343044, and is detected at IR, Optical and X-ray wavelengths. 
%(WISEA~J132105.56+341500.9 and 2CXO~J132105.4+341459)
%It is a radio-loud AGN with a spectroscopic redshift of $z\text{ = } 0.452$ \citep{Caccianiga2008}, and is included in the quasars and active nuclei catalogue of \citet{Veron2010}. 

\begin{longtable}{lll}
\caption{The main properties of the radio source NVSS~132105+341502 (see section \ref{3.1.5})}
\label{T3} \\
\makebox[\textwidth]{
    \begin{tabular}{lccccc}
        \toprule
         \multirow{3}{*}{Source name} & \multirow{3}{*}{Type } & \multirow{3}{*}{$z$ } & \multirow{3}{*}{S$_{1.4~\rm{GHz}}$ (mJy)} & \multirow{3}{*}{Size (") } & \multirow{3}{*}{Other ID }  \\
        & & & &  \\
        \midrule
        NVSS~132105+341502 & Radio-loud AGN & 0.452 & $474\pm17$ & $< 17$ & 87GB~131847.5+343057 \\
        & & & & & B2~1318+34B \\  
        & & & & & 6C B131847.2+343044 \\
        & & & & & WISEA~J132105.56+341500.9 \\
        & & & & & 2CXO~J132105.4+341459 \\
        \midrule
        \bottomrule
    \end{tabular}}
    \end{longtable}

\vspace{1cm}

%nearby low surface brightness galaxy MCG+08-36-003 (with further cross-identifications including CGCG 257-007) with $M_{V}\text{ = }14.0$ and a spectroscopic redshift $z\text{ = }0.02595$. In the optical, MCG+08-36-003 straddles the 4.2 arcminute search cone, and appears to be part of a group of galaxies that includes UGC 11465 ($z\text{ = }0.026061$), WISEA J194140.93+503832.5 ($z\text{ = }0.027592$) and 2MASX J19414453+5037151 ($z\text{ = }0.027805$).     

\begin{longtable}{lll}
\caption{Details of the galaxy group associated with 3C402 (see section \ref{3.1.5})}
\label{T4} \\
\makebox[\textwidth]{
    \begin{tabular}{lclc}
        \toprule
         \multirow{3}{*}{Source name} & \multirow{3}{*}{Type } & \multirow{3}{*}{$z$ } & \multirow{3}{*}{Other ID }  \\
        & &  &  \\
        \midrule
        MCG+08-36-003 & galaxy & 0.02595 & CGCG 257-007 \\
        UGC 11465 & galaxy & 0.026061 & - \\ 
        WISEA J194140.93+503832.5 & galaxy &  0.027592 & - \\
        2MASX J19414453+5037151 & galaxy & 0.027805 & - \\
        \midrule
        \bottomrule
    \end{tabular}}
    \end{longtable}
    
\vspace{1cm}

\begin{longtable}{lll}
\caption{CWTFM and prevalence figures for extraterrestrial transmitters of astrophysical exotica}
\label{T5} \\

\makebox[\textwidth]{
    \begin{tabular}{lllccccccc}
        \toprule
        
        \multirow{3}{*}{BL Field name} & \multirow{3}{*}{Source name} & \multirow{3}{*}{Type } & \multirow{3}{*}{$d_{l}$ (Mpc) } & \multirow{3}{*}{Offset ($\arcmin$)} & \multirow{3}{*}{Beam  }&  \multirow{3}{*}{EIRP$_{\rm min}$} & \multirow{3}{*}{$N_{*}$} & \multirow{3}{*}{$log$TR} &\multirow{3}{*}{CWTFM} \\
        & & & & & & & \\
        & & & & & Response (\% ) & & \\
        \midrule
        BL~J033901.1-053736.5 & WISEA J033845.71-053842.1 & Galaxy & 46.7 & 3.98 & 53.6  & $9.8 \times 10^{23}$ &  $\sim 10^{11\pm0.3}$ & $-10.6\pm0.3$& 919\\
        BL J143328.4+525435.3 & WHL~J143313.0+525748 + others & Cluster & 205.3  & 3.96 & 54.0  & $1.9 \times 10^{25}$ & $\sim 10^{12\pm0.5}$ & $-11.6\pm0.5$ & 1781 \\
        BL J233623.1+020609.0 & NGC 7714 \& NGC 7715 & Interacting system & 40.3 & 3.88 & 55.3  & $7.5 \times 10^{23}$ & $\sim 2 \times 10^{11\pm0.3}$ & $-10.9\pm0.3$ & 352 \\
        BL J080812.8+210612.2 & WISEA~J080816.59+210857.0 & AGN & 672.2 & 2.88 & 72.2  & $2.7 \times 10^{26}$ & $\sim 10^{11\pm0.3}$ & $-10.6\pm0.3$ & 253125 \\
        BL J194148.6+503127.5 & 3C402 + others  & Galaxy Group & 113.4  & 4.2 & 50.0  & $5.4 \times 10^{24}$ & $\sim 4 \times 10^{11\pm0.3}$ & $-11.2\pm0.3$ & 1265 \\
        BL J184244.0+593803.8 & MM184222+593828 & Lens system & 35099.1  & 2.78 & 73.8  & $7.6 \times 10^{29}\mu^{-1}$ & $\sim 10^{11\pm0.3}$ & $-10.6\pm0.3$ & $712500000\mu^{-1}$ \\
        \midrule
        \bottomrule
    \end{tabular}}
    \end{longtable}
\end{center}

\label{lastpage}
\end{document}